\documentstyle[12pt,amsmath]{article}

\newcommand{\be}{\begin{equation}}
\newcommand{\ee}{\end{equation}}
\newcommand{\ba}{\begin{eqnarray}}
\newcommand{\ea}{\end{eqnarray}}
\newcommand{\p}{\partial}

\begin{document}

\begin{titlepage}
\vspace{1cm}

\begin{centering}

{\Large \bf Quantum complex scalar fields \\ \bigskip and  noncommutativity}

\vspace{.5cm}
\vspace{1cm}

{\large Ricardo Amorim$^{a}$} and 
{\large Everton M. C. Abreu$^{b}$}

\vspace{0.5cm}

$^a$Instituto de F\'{\i}sica, Universidade Federal
do Rio de Janeiro,\\
Caixa Postal 68528, 21945-970,  Rio de Janeiro, Brazil\\[1.5ex]

$^b$Grupo de F\' isica Te\'orica e Matem\'atica F\' isica,
Departamento de F\'{\i}sica,\\
 Universidade Federal Rural do Rio de Janeiro,\\
BR 465-07, 23890-000, Serop\'edica, Rio de Janeiro, Brazil\\

\bigskip

\today

\begin{abstract}
In this work we analyze complex scalar fields using a new framework where the object of noncommutativity 
$\theta^{\mu\nu}$ represents independent degrees of freedom. In a first quantized formalism, $\theta^{\mu\nu}$ 
and its canonical momentum $\pi_{\mu\nu}$ are seen as operators  living in some Hilbert space.
This structure is compatible with the minimal canonical extension of the Doplicher-Fredenhagen-Roberts (DFR) algebra
and is invariant under an extended Poincar\'e group of symmetry. 
In a second quantized formalism perspective, we present an explicit form for the extended Poincar\'e
generators and the same algebra is generated via generalized Heisenberg relations.  We also introduce a source term and construct the general solution for the complex scalar fields using the
Green's function technique.
\end{abstract}

\end{centering}

\vspace{1cm}


\vfill

\noindent$^a${\tt amorim@if.ufrj.br}\\
$^b$\noindent{\tt evertonabreu@ufrrj.br}

\end{titlepage}

\pagebreak

\section{Introduction}
\renewcommand{\theequation}{\arabic{equation}}
\setcounter{equation}{0}

\bigskip 


Through the last years  space-time noncommutativity has been a target of intense analysis.
After the first published work by Snyder \cite{Snyder} a huge amount of papers has appeared in the literature.
The connection with strings \cite{Strings}, gravity \cite{QG,Hull,SW} and noncommutative field theories (NCFT) \cite{NCFT} 
brought attention to the subject.  

The fundamental  idea is that  space-time may loose its standard properties at very high energy regimes. One of the  approaches to study space-time at those regimes could be related with the noncommutativity of the coordinates \cite{Hull,SW}.  Other approaches to noncommutativity can also be given in a global way, generalizing
some of the celebrated Connes ideas \cite{connes}.

Most of the  theories cited above  pinpoint to the fact that  at 
Planck scale, the space-time coordinates $x^\mu$ have to be replaced by Hermitean operators $\mathbf{x}^\mu$
 obeying the commutation relations
\begin{equation}
\label{02}
[{\mathbf x}^\mu,{\mathbf x}^\nu] = i {\mathbf \theta}^{\mu\nu}\,\,,
\end{equation} 

\noindent   where ${\mathbf \theta}^{\mu\nu}$ is considered as a constant antissimetric matrix.
 Although it maintains translational invariance,  
Lorentz symmetry is broken \cite{NCFT} or correspondingly the rotation symmetry for non-relativistic
theories. 
The violation of Lorentz invariance is problematic, among other facts, because it brings effects such as vacuum birefringence \cite{Jaeckel}. 
Other approaches are possible 
\cite{Carlson,Haghighat,Carone,Ettefaghi,Morita,Saxell},
 permitting 
to construct Lorentz invariant theories, by considering in some sense ${\mathbf \theta}^{\mu\nu}$ as independent degrees of freedom.
These approaches are related to seminal works
 by Doplicher, Fredenhagen and Roberts (DFR) \cite{DFR}, which contain a blend
of the principles of classical General Relativity and Quantum Mechanics. 
 Their theory
constrains localizability in a quantum spacetime, that has to be extended to include $\theta^{\mu\nu}$ 
as an independent set of coordinates. 

The structure of the DFR theory, besides equation (\ref{02}) comprises
\begin{equation}
\label{03}
[{\mathbf x}^\mu,{\mathbf \theta}^{\alpha\beta}] =0 \qquad \mbox{and} \qquad
[{\mathbf \theta}^{\mu\nu},{\mathbf \theta}^{\alpha\beta}]=0\,\,,
\end{equation}

\noindent with subsidiary  quantum conditions
\begin{equation}
\label{04}
{\mathbf \theta}_{\mu\nu}{\mathbf \theta}^{\mu\nu} =0 \qquad \mbox{and} \qquad
({1\over4}{}^{*}{\mathbf \theta}^{\mu\nu}{\mathbf \theta}_{\mu\nu})^2=\lambda_P^8\,\,,
\end{equation}

\noindent where ${}^{*}{\mathbf \theta}_{\mu\nu}={1\over2}\epsilon_{\mu\nu\rho\sigma}{\mathbf
\theta}^{\rho\sigma}$and $\lambda_P$ is the Planck length. 

 The main motivation of DFR to
study the relations (\ref{02}) and (\ref{03}) was the belief that a tentative of exact measurements involving space-time localization could confine
photons due to gravitational fields.  This 
phenomenon is directly related to (\ref{02}) and (\ref{03}) together with (\ref{04}). In a somehow different perspective, other
 relevant results are obtained in 
\cite{Carlson,Haghighat,Carone,Ettefaghi,Morita,Saxell} relying on conditions ($\ref{02}$) and
($\ref{03}$).  These 
authors use the value  of $\theta$ taken as a mean with some weigh function,
generating Lorentz invariant theories and providing a connection with usual theories constructed in an ordinary $D=4$ space-time.

In a recent series of works \cite{Amorim1,Amorim4,Amorim5,Amorim2} a new version of noncommutative
quantum mechanics (NCQM) has been presented by one of us, where not only
the coordinates ${\mathbf x}^\mu$ and their canonical momenta ${\mathbf p}_\mu$ are considered as operators in a Hilbert space
${\cal H}$, but also the objects of noncommutativity $\theta^{\mu\nu}$ and their canonical conjugate momenta $\pi_{\mu\nu}$.
All these operators belong to the same algebra and have the same hierarchical level, introducing  a minimal canonical
extension of the DFR algebra. This enlargement of the usual
set of Hilbert space operators allows the theory to be invariant under the rotation group $SO(D)$, as showed 
in detail in Ref. \cite{Amorim1,Amorim2}, when the treatment is a nonrelativistic one.  Rotation invariance in a nonrelativistic theory, is fundamental if one intends to
describe any physical system in a consistent way. In Ref. \cite{Amorim4,Amorim5}, the corresponding relativistic treatment is presented, which permits to implement Poincar\'{e} invariance as
a dynamical symmetry   \cite{Iorio} in  NCQM  \cite{NCQM}. 
In the present work we essentially consider the "second quantization" of the model discussed in Ref \cite{Amorim4}, showing that the extended Poincar\'e
symmetry here is generated via generalized Heisenberg relations, giving the same algebra desplayed in \cite{Amorim4,Amorim5}.

The structure of this paper is organized as: after  this introductory section, the minimal canonical extension of the DFR algebra is reviewed  in Section {\bf 2}.  In Section {\bf 3}, we 
introduce the new noncommutative charged Klein-Gordon theory in this $D=10$,  $x+\theta$ space and 
analyze its symmetry structure, associated with the invariance of the action under some extended Poincar\'e (${\cal P}'$) group.  This symmetry structure is also displayed at the second quantization level, constructed via generalized Heisenberg relations. In Section {\bf 4}
we expand the fields in a plane wave basis in order to solve the equations of motion
using the Green's functions formalism adapted for this new $(x+\theta)\: D=4+6$ space.
The conclusions and perspectives are reserved to  the last section.

\section{The minimal canonical extension of the DFR algebra}

Besides (\ref{02}), (\ref{03}), the minimal canonical extension \cite{Amorim4} of the DFR algebra\cite{DFR} is given by
\begin{eqnarray}
\label{i2}
& & [{\mathbf x}^\mu,{\mathbf p}_\nu ] = i\delta^{\mu}_{\nu}\,\,,\nonumber\\
& & [{\mathbf p}_\mu,{\mathbf p}_\nu ] = 0\,\,,\nonumber\\
& & [{\mathbf \theta}^{\mu\nu},{\mathbf \pi}_{\rho\sigma}]=i\delta^{\mu\nu}_{\,\,\,\rho\sigma}\,\,, \nonumber\\
& & [{\mathbf \pi}_{\mu\nu},{\mathbf \pi}_{\rho\sigma}]=0\,\,,\nonumber\\
& & [{\mathbf p}_\mu,{\mathbf \theta}^{\rho\sigma}]=0\,\,,\nonumber\\
& & [{\mathbf p}_\mu,{\mathbf \pi}_{\rho\sigma}]=0\,\,,\nonumber\\
& & [{\mathbf x}^\mu,{\mathbf \pi}_{\rho\sigma}]=-{i\over2}\delta^{\mu\nu}_{\,\,\,\rho\sigma}p_\nu\,\,,
\end{eqnarray}

\noindent where
$\delta^{\mu\nu}_{\,\,\,\,\rho\sigma}=\delta^\mu_\rho\delta^\nu_\sigma-\delta^\mu_\sigma\delta^\nu_\rho$.
The relations above are consistent under all possible Jacobi identities. 
Notice that the ordinary  form of the  Lorentz generator,
given by ${ \mathbf l}^{\mu\nu}= { \mathbf x}^\mu{\mathbf p}^\nu-{\mathbf
x}^\nu{\mathbf p}^\mu$, fails to close in an algebra if  (\ref{02}) is adopted, even if one considers ${\mathbf \theta}^{\mu\nu}$ as constant.
\bigskip

The relations (\ref{02}), (\ref{03}) and (\ref{i2}) allows us to utilize \cite{Gracia},
\begin{equation}
\label{i16}
{ \mathbf M}^{\mu\nu}= { \mathbf X}^\mu{\mathbf p}^\nu-{\mathbf X}^\nu{\mathbf
p}^\mu-{\mathbf \theta}^{\mu\sigma}{\mathbf \pi}_\sigma^{\,\,\nu}+{\mathbf
\theta}^{\nu\sigma}{\mathbf \pi}_\sigma^{\,\,\mu}
\end{equation}

\noindent  as the  generator of the Lorentz group, where   \cite{NCQM}
\begin{equation}
\label{i5}
{\mathbf X}^\mu={\mathbf x}^\mu+{1\over2}{\mathbf\theta}^{\mu\nu}{\mathbf p}_\nu\,\,,
\end{equation}

\noindent and we see that the proper algebra is closed, i.e.,
\begin{equation}
\label{i17}
[{\mathbf M}^{\mu\nu},{\mathbf M}^{\rho\sigma}]=i\eta^{\mu\sigma}{\mathbf
M}^{\rho\nu}-i\eta^{\nu\sigma}{\mathbf M}^{\rho\mu}-i\eta^{\mu\rho}{\mathbf
M}^{\sigma\nu}+i\eta^{\nu\rho}{\mathbf M}^{\sigma\mu}\,\,.
\end{equation}

\noindent Now $\mathbf M^{\mu\nu}$  
generates the expected symmetry transformations when acting on all the operators in
Hilbert space.  
Namely, by defining the dynamical transformation of an arbitrary operator ${\mathbf A}$ in
$\cal H$ in such a way that
$\delta {\mathbf A}=i[ {\mathbf A}, {\mathbf G}]$, where 
\be
{\mathbf G}={1\over2}\omega_{\mu\nu}{\mathbf M}^{\mu\nu}-a^\mu{\mathbf
p}_\mu+{1\over2}b^{\mu\nu}{\mathbf \pi}_{\mu\nu}\,\,, 
\ee
and $\omega^{\mu\nu}=-\omega^{\nu\mu}$, $a^\mu$, $b^{\mu\nu}=-b^{\nu\mu}$ are
infinitesimal parameters, it follows that

\begin{subequations}
\label{i19}
\ba 
\delta {\mathbf x}^\mu &=& \omega ^\mu_{\,\,\,\,\nu}{\mathbf x}^\nu+a^\mu+{1\over2}b^{\mu\nu}p_\nu\,\,,\label{i19a} \\ 
\delta {\mathbf X}^\mu &=& \omega ^\mu_{\,\,\,\,\nu}{\mathbf X}^\nu+a^\mu\,\,,\label{i19b} \\ 
\delta{\mathbf p}_\mu  &=& \omega _\mu^{\,\,\,\,\nu}{\mathbf p}_\nu\,\,, \label{i19c} \\
\delta{\mathbf \theta}^{\mu\nu} &=& \omega ^\mu_{\,\,\,\,\rho}{\mathbf
\theta}^{\rho\nu}+ \omega ^\nu_{\,\,\,\,\rho}{\mathbf \theta}^{\mu\rho}+b^{\mu\nu}\,\,,\label{i19d} \\
\delta{\mathbf \pi}_{\mu\nu} &=& \omega _\mu^{\,\,\,\,\rho}{\mathbf \pi}_{\rho\nu}+
\omega _\nu^{\,\,\,\,\rho}{\mathbf \pi}_{\mu\rho}\,\,, \label{i19e} \\
\delta {\mathbf M}^{\mu\nu} &=& \omega ^\mu_{\,\,\,\,\rho}{\mathbf M}^{\rho\nu}+
\omega ^\nu_{\,\,\,\,\rho}{\mathbf M}^{\mu\rho}+a^\mu{\mathbf p}^\nu-a^\nu{\mathbf
p}^\mu+b^{\mu\rho}{\mathbf \pi}_\rho^{\,\,\,\,\nu}+ b^{\nu\rho}{\mathbf
\pi}_{\,\,\,\rho}^{\mu}\,\,,\label{i19f}
\ea
\end{subequations}

\noindent  generalizing the action of the Poincar\'{e} group ${\cal P}$ in order to include
$\theta$ and $\pi$ transformations. Let us refer to this group as ${\cal P}'$. The ${\cal P}'$ transformations 
close in an algebra. Actually, 

\begin{equation}\label{xxx}
[\delta_2,\delta_1]\,{\mathbf A}=\delta_3\,{\mathbf A}\,\,,
\end{equation}

\noindent and the parameters composition rule is given by
\begin{eqnarray}
\label{i19bb}
& & \omega^\mu_{3\,\,\,\,\nu}=\omega^\mu_{1\,\,\,\,\alpha}\omega^\alpha_{2\,\,\,\,\nu}-\omega^\mu_{2\,\,\,\,\alpha}\omega^\alpha_{1\,\,\,\,\nu}\,\,,
\nonumber\\
& & a_3^\mu=\omega^\mu_{1\,\,\,\nu}a_2^\nu-\omega^\mu_{2\,\,\,\nu}a_1^\nu\,\,,\nonumber\\
& & b_3^{\mu\nu}=\omega^\mu_{1\,\,\,\rho}b_2^{\rho\nu}-\omega^\mu_{2\,\,\,\rho}b_1^{\rho\nu}-\omega^\nu_{1\,\,\,\rho}b_2^{\rho\mu}+
\omega^\nu_{2\,\,\,\rho}b_1^{\rho\mu}\,\,.
\end{eqnarray}
The symmetry structure displayed in (\ref{i19}) is discussed in details in \cite{Amorim4}.

Also in \cite{Amorim4}, it was studied how these symmetries could be dynamically implemented in a Lagrangian formalism.
Theories that are invariant under ${\cal P}$ and ${\cal P}'$ were considered. The underlying
point relies in the use of the Noether's formalism adapted to such $x+\theta$
extended space.  Moreover, this last cited work introduced possible NCQM actions constructed with
the Casimir operators of ${\cal P}'$. As can be verified, if we define ${\mathbf M}_1^{\mu\nu}={ \mathbf X}^\mu{\mathbf p}^\nu-{\mathbf X}^\nu{\mathbf p}^\mu$ and ${\mathbf M}_2^{\mu\nu}=-{\mathbf \theta}^{\mu\sigma}{\mathbf \pi}_\sigma^{\,\,\nu}+{\mathbf
\theta}^{\nu\sigma}{\mathbf \pi}_\sigma^{\,\,\mu}$, both satisfying (\ref{i17}), one can verify that four Casimir operators for 
${\cal P}'$ can be constructed. Namely, the first two of such invariant operators are the usual one given by
 ${\mathbf C}_1={\mathbf p}^2$ and ${\mathbf C}_2={\mathbf s}^2$, where 
${\mathbf s}_\mu={1\over2}\epsilon_{\mu\nu\rho\sigma}{\mathbf M}_1^{\nu\rho}{\mathbf p}^\sigma$  is the Pauli-Lubanski vector. 
The last two are defined as  ${\mathbf C}_3={\mathbf \pi}^2$ and ${\mathbf C}_4={\mathbf M}_2^{\mu\nu}{\mathbf \pi}_{\mu\nu}$.
The important point to be stressed here is that the usual  Casimir operators for the Poincar\'e group are kept by the theory, which does not destroy the usual classification scheme for the elementary particles. The same scheme can also be extended to fermionic fields \cite{Amorim5}. Furthermore, it was shown that a corresponding classical underlying theory can also be constructed, as the one given in 
Ref. \cite{Amorim2}.

\section{The action and symmetry relations}

An important point is that, due to (\ref{02}), the operator ${\mathbf x}^\mu$ can not
be used to label a possible basis in ${\cal H}$. However, as the components of
${\mathbf X}^\mu$ commute, as can be verified from (\ref{i2}) and  (\ref{i5}), their
eigenvalues  can be used for such purpose. From now on let us denote  by $x$ and
$\theta$ the eigenvalues of ${\mathbf X}$ and ${\mathbf\theta}$. In  \cite{Amorim4}
we have considered these points with some detail and have proposed a way for
constructing  actions representing possible field theories in this extended
$x+\theta$ space-time.  One of such actions, generalized in order to permit the
scalar fields to be complex, is given by

\begin{equation}
\label{b8}
S=-\int d^{4}\,x\,d^{6}\theta\,\, \Big\{\,\partial^\mu\phi^*\partial_\mu\phi +
\,{{\lambda^2}\over4}\,\partial^{\mu\nu}\phi^*\partial_{\mu\nu}\phi  
+m^2\,\phi^*\phi\Big\}\,\,,
\end{equation}

\noindent where $\lambda$ is a parameter with dimension of length, as the Planck
length, which is introduced due to dimensional reasons. Here we are also suppressing
a possible  factor $\Omega(\theta)$ in the measure, which is a scalar weight
function, used in Refs.   
\cite{Carlson}-\cite{Saxell}, 
in a noncommutative gauge
theory context,  to make the connection between the $D=4+6$
and the $D=4$ formalisms. Also $\Box= \partial^\mu\partial_\mu  $, with
$\partial_\mu={{\partial}\over{\partial {x}^\mu}}$ and 
$\Box_{\theta}={1\over2}\partial^{\mu\nu}\partial_{\mu\nu}$,
where $\partial_{\mu\nu}={{\partial\,\,\,}\over{\partial {\theta}^{\mu\nu}}}\,\,$ and  
$\eta^{\mu\nu}=diag(-1,1,1,1)$.
\bigskip

The corresponding Euler-Lagrange equation reads

\begin{eqnarray}
\label{c6}
{{\delta S}\over{\delta\phi}}& =& \,(\Box +\lambda^2\Box_\theta- m^2)\phi^* \nonumber\\
& =& \,\,\,0\,\,,
\end{eqnarray}

\noindent with a similar equation of motion for $\phi$. The action (\ref{b8}) is
invariant under the transformation 

\begin{equation}
\label{i19cc}
\delta
\phi=-(a^\mu+\omega^\mu_{\,\,\,\nu}x^\nu)\,\partial_\mu\phi-{1\over2}(b^{\mu\nu}+2\omega^\mu_{\,\,\,\rho}\theta^{\rho\nu})\,\partial_{\mu\nu}\phi\,\,,
\end{equation}

\noindent besides the phase transformation
\begin{equation}
\label{i19dd}
\delta \phi=-i\alpha\,\phi\,\,,
\end{equation}

\noindent with  similar expressions for $\phi^*$, obtained from (\ref{i19cc}) and
(\ref{i19dd}) by complex conjugation. 
 We observe that 
\noindent (\ref{i19c}) closes in an algebra, as in (\ref{xxx}), with the same
composition rule defined in (\ref{i19bb}). That equation defines  how a complex
scalar field transforms in the $x+\theta$ space under ${\cal P}'$. The
transformation subalgebra generated by (\ref{i19d}) is of course Abelian, although
it could be directly generalized to a more general setting.

Associated with those symmetry transformations, we can define the conserved
currents \cite{Amorim4} 

\begin{eqnarray}
\label{92}
& & j^\mu={{\partial {\cal L}}\over{\partial\partial_\mu\phi}}\delta\phi+
\delta\phi^*{{\partial {\cal L}}\over{\partial\partial_\mu\phi^*}}+
{\cal L}\delta x^\mu\,\,,\nonumber\\
& & j^{\mu\nu}={{\partial {\cal L}}\over{\partial\partial_{\mu\nu}\phi}}\delta\phi+
\delta\phi^*{{\partial {\cal L}}\over{\partial\partial_{\mu\nu}\phi^*}}
+{\cal L}\delta \theta^{\mu\nu}\,\,.
\end{eqnarray}

\noindent Actually, by using (\ref{i19c}) and (\ref{i19d}),  as well as (\ref{i19}b)
and (\ref{i19}d), we can show, after some algebra, that 

\begin{equation}\label{94}
\partial_\mu j^\mu+\partial_{\mu\nu}j^{\mu\nu}=-{{\delta
S}\over{\delta\phi}}\delta\phi-\delta\phi^*{{\delta S}\over{\delta\phi^*}}\,\,.
\end{equation}

Similar calculations can be found, for instance, in \cite{Amorim4}.
The expressions above allow us to derive a specific charge 

\begin{equation}
\label{95}
Q=-\int d^3 x d^6 \theta\, j^0\,\,,
\end{equation}

\noindent  for each kind of conserved symmetry encoded in (\ref{i19c}) and
(\ref{i19d}), since

\begin{equation}
\label{96}
\dot Q=\int d^3 x d^6 \theta\, (\partial_i j^i+{1\over2}\partial_{\mu\nu}j^{\mu\nu}) 
\end{equation}

\noindent vanishes as a consequence of the divergence theorem in this $x+\theta$ extended space. Let us consider each
specific symmetry  in (\ref{i19c}) and (\ref{i19d}). For usual $x$-translations, we can write
$j^0=j^0_\mu a^\mu$, permitting to define the total momentum

\begin{eqnarray}
\label{95.1}
P_\mu& =& -\int d^3 x d^6 \theta\, j^0_\mu\nonumber\\
& =& \int d^3 x d^6
\theta\,(\dot\phi^{*}\partial_\mu\phi+\dot\phi\partial_\mu\phi^{*}-{\cal
L}\delta^0_\mu)\,\,.
\end{eqnarray}

\noindent For $\theta$-translations, we can write that $j^0=j^0_{\mu\nu}b^{\mu\nu}$,
giving
\begin{eqnarray}
\label{96.1}
P_{\mu\nu}& =& -\int d^3 x d^6 \theta\, j^0_{\mu\nu}\nonumber\\
& =& {1\over2}\int d^3 x d^6
\theta\,(\dot\phi^*\partial_{\mu\nu}\phi+\dot\phi\partial_{\mu\nu}\phi^*)\,\,.
\end{eqnarray}
 
\noindent In a similar way we define the Lorentz charge. By using the operator
\begin{equation}
\label{97}
\Delta_{\mu\nu}=x_{[\mu}\partial_{\nu]}+\theta_{[\mu}^{\,\,\,\alpha}\partial_{\nu]\alpha}\,\,,
\end{equation}

\noindent and defining $j^0={\bar j}^0_{\mu\nu}\omega^{\mu\nu}$, we can write
\begin{eqnarray}
\label{98}
M_{\mu\nu}& =& -\int d^3 x d^6 \theta\, {\bar j}^0_{\mu\nu}\nonumber\\
& =& \int d^3 x d^6
\theta\,(\dot\phi^*\Delta_{\nu\mu}\phi+\dot\phi\Delta_{\nu\mu}\phi^*-{\cal
L}\delta^0_{[\mu}x_{\nu]})\,\,.
\end{eqnarray}

 At last, for the symmetry given by (\ref{i19d}), we get the $U(1)$ charge as
\begin{equation}
\label{99}
{\cal Q}= i\,\int d^3 x d^6 \theta\,(\dot\phi^*\phi-\dot\phi\phi^*)\,\,.
\end{equation}

Now let us show that these charges generate the appropriate field transformations 
(and dynamics) in a quantum scenario, as generalized Heisenberg relations.
To start the quantization of such theory, we can define as usual the field momenta
\begin{eqnarray}
\label{100}
& & \pi={{\partial{\cal L}}\over{\partial\dot\phi}}=\dot\phi^*\,\,,\nonumber\\
& & \pi^*={{\partial{\cal L}}\over{\partial\dot\phi^*}}=\dot\phi\,\,,
\end{eqnarray}

\noindent satisfying the non vanishing equal time commutators ( in what follows the
commutators are to be understood as equal time commutators )
\begin{eqnarray}
\label{101}
& & [\pi(x,\theta),\phi(x',\theta')]=-i\delta^3(x-x')\delta^6(\theta-\theta')\,\,,\nonumber\\
& & [\pi^*(x,\theta),\phi^*(x',\theta')]=-i\delta^3(x-x')\delta^6(\theta-\theta')\,\,.
\end{eqnarray}

The strategy now is just to generalize the usual field theory and rewrite the
charges (\ref{95.1}-\ref{99}) by eliminating the time derivatives of the fields in
favor of the field momenta. After that we use (\ref{101}) to dynamically generate
the symmetry operations. In this spirit, accordingly to (\ref{95.1}) and (\ref{100}),
the spatial translation is generated by
\begin{equation}
\label{102}
P_i=\int d^3 x d^6 \theta\,(\pi\partial_i\phi+\pi^*\partial_i\phi^*)\,\,,
\end{equation}

\noindent and it is trivial to verify, by using (\ref{101}), that 
\begin{equation}
\label{103}
[P_i,{\cal Y}(x,\theta)]=-i\partial_i{\cal Y}(x,\theta)\,\,,
\end{equation}

\noindent where ${\cal Y}$ represents $\phi,\,\phi^*,\pi$ or $\pi^*$. 
The dynamics is generated by 
\begin{equation}
\label{102.1}
P_0=\int d^3 x d^6
\theta\,(\pi^*\pi+\partial^i\phi^*\partial^i\phi+{{\lambda^2}\over4}\partial^{\mu\nu}\phi^*\partial_{\mu\nu}\phi+m^2\phi^*\phi)
\end{equation}

\noindent accordingly to (\ref{95.1}) and (\ref{100}). At this stage it is convenient to
assume that classically $\partial^{\mu\nu}\phi^*\partial_{\mu\nu}\phi\geq0$ to
assure that the Hamiltonian $H=P_0$ is positive definite.  By using the fundamental
commutators (\ref{101}), the equations of motion (\ref{c6}) and the definitions
(\ref{100}), it is possible to prove the Heisenberg relation
\begin{equation}
\label{103.1}
[P_0,{\cal Y}(x,\theta)]=-i\partial_0{\cal Y}(x,\theta)\,\,.
\end{equation}

The $\theta$-translations, accordingly to (\ref{96.1}) and (\ref{100}), are generated by
\begin{equation}
\label{104}
P_{\mu\nu}=\int d^3 x d^6
\theta\,(\pi\partial_{\mu\nu}\phi+\pi^*\partial_{\mu\nu}\phi^*)\,\,,
\end{equation}

\noindent and one obtains trivially  by (\ref{101}) that
\begin{equation}
\label{105}
[P_{\mu\nu},{\cal Y}(x,\theta)]=-i\partial_{\mu\nu}{\cal Y}(x,\theta)\,\,.
\end{equation}

Lorentz transformations are generated by (\ref{98}) in a similar way. The spatial
rotations generator is given by
\begin{equation}
\label{106}
M_{ij}=\int d^3 x d^6 \theta\,(\pi\Delta_{ji}\phi+\pi^*\Delta_{ji}\phi^*)\,\,,
\end{equation}

\noindent  while the boosts are generated by
\begin{eqnarray}
\label{107}
M_{0i}& =& {1\over2}\int d^3 x d^6 \theta\,\Big\{\pi^*\pi
x_i-x_0(\pi\partial_i\phi+\pi^*\partial_i\phi^*)\nonumber\\
& +& \pi(2\theta_{[i}^{\,\,\,\gamma}\partial_{0]\gamma}-x_0\partial_i)\phi+\pi^*(2\theta_{[i}^{\,\,\,\gamma}\partial_{0]\gamma}-x_0\partial_i)\phi^*\nonumber\\
& +& (\partial_j\phi^*\partial_j\phi+{{\lambda^2}\over{4}}\partial^{\mu\nu}\phi^*\partial_{\mu\nu}\phi+m^2\phi^*\phi)x_i\Big\}\,\,.
\end{eqnarray}

\noindent As can be verified in a direct way for (\ref{106}) and in a little more
indirect way for (\ref{107}),
\begin{equation}
\label{108}
[M_{\mu\nu},{\cal Y}(x,\theta)]=i\Delta_{\mu\nu}{\cal Y}(x,\theta)\,\,,
\end{equation}

\noindent for any dynamical quantity ${\cal Y}$, where $\Delta_{\mu\nu}$ has been
defined in (\ref{97}). At last
we can rewrite (\ref{99}) as

\begin{equation}
\label{109}
{\cal Q}= i\,\int d^3 x d^6 \theta\,(\pi\phi-\pi^*\phi^*)\,\,,
\end{equation}

\noindent generating (\ref{i19d}) and its conjugate, and similar expressions for
$\pi$ and $\pi^*$. So, the $\cal{P}$' and (global) gauge transformations can be generated by the action of the operator

\begin{equation}
\label{110}
{G}={1\over2}\omega_{\mu\nu}{M}^{\mu\nu}-a^\mu{P}_\mu+{1\over2}b^{\mu\nu}{P}_{\mu\nu}-\alpha {\cal Q}
\end{equation}

\noindent over the complex fields and their momenta, by using the canonical commutation relations (\ref{101}). In this way the $\cal{P}$' and gauge transformations are generated as generalized Heisenberg relations. This is a new result that shows the consistence of the above formalism. Furthermore, there are also four Casimir operators defined with the operators given above, with the same form as those previously defined at a first quantized perspective.
So, the structure displayed above is very similar to the usual one found
in ordinary quantum complex scalar fields. We can go one step further,
by expanding the fields and momenta in modes, giving as well some order prescription,  to define the
relevant Fock space, spectrum, Green's functions and all the basic structure related to free bosonic
fields. In what follows we consider some of these issues and postpone others for a forthcoming work \cite{aa}.

\section{Plane waves and Green's functions}

In order to explore a litle more the framework described in the last sections, 
 let us rewrite the generalized charged Klein-Gordon action (\ref{b8}) with source terms as

\be
\label{0.0}
S\,=\,-\,\int d^4 x d^6 \theta \Big\{\p^\mu\phi^* \p_\mu \phi \,+\,{\lambda^2 \over 4}\,\p^{\mu\nu}\phi^* \p_{\mu\nu} \phi \,+\,m^2\,\phi^* \phi \,+\,J^* \phi\,+\,J \phi^* \Big\}\,\,.
\ee

\noindent The corresponding equations of motion   are

\be
\label{0.1}
(\Box +\lambda^2\Box_\theta- m^2)\phi(x,\theta)\,=\,J(x,\theta)
\ee

\noindent as well as its complex conjugate one.
 We have the following formal solution

\be
\label{0.2.1}
\phi(x,\theta)\,=\,\phi_{J=0}(x,\theta)\,+\,\phi_J (x,\theta)\,
\ee

\noindent where, clearly, $\phi_{J=0}(x,\theta)$ is the source free solution 
and $\phi_J (x,\theta)$ is the solution with $J\neq 0$. 

The Green's function for
(\ref{0.1}) satisfies

\be
\label{0.3}
(\Box +\lambda^2\Box_\theta- m^2)\,G(x-x',\theta-\theta')=\,\delta^4 (x-x')\,\delta^6 (\theta-\theta')\,\,,
\ee

\noindent where $\delta^4 (x-x')$ and $\delta^6 (\theta-\theta')$ are the Dirac's delta functions

\be
\label{0.9}
\delta^4 (x-x')\,=\,{1\over (2\pi)^4}\,\int d^4 K_{(1)} \,e^{iK_{(1)}\cdot (x-x')}\,\,,
\ee
\be
\label {0.10}
\delta^6 (\theta-\theta')\,=\,{1\over (2\pi)^6}\,\int d^6 K_{(2)} \,e^{iK_{(2)}\cdot (\theta-\theta')}\,\,.
\ee

Now let us define

\be
\label{0.7}
X\,=\,(x^\mu,{1\over \lambda}\,\theta^{\mu\nu})
\ee
\noindent and
\be
\label{0.8}
K\,=\,(K^\mu_{(1)},\lambda\,K^{\mu\nu}_{(2)})\,\,,
\ee

\noindent where $\lambda$ is a parameter that carries the dimension of length, as said before.  From (\ref{0.7}) and (\ref{0.8}) we 
write that $K\cdot X = K_{(1)\mu}\,x^\mu + {1\over 2} K_{(2)\mu\nu}\,\theta^{\mu\nu}$.  The factor ${1\over 2}$
is introduced in order to eliminate repeated terms. In what follows it will also be considered that 
$d^{10}K=d^4 K_{(1)}d^6 K_{(2)}$ and $d^{10}X=d^4 x\,d^6 \theta$.

So, from (\ref{0.1}) and (\ref{0.3}) we formally have that
\be
\label{0.4}
\phi_J (X)\,=\,\int d^{10} X' G(X-X')\,J(X')\,\,.
\ee

To derive an explicit form for the Green's function, let us expand $G(X-X')$ in terms of plane waves. Hence, we can write that,

\be
\label{0.12}
G(X-X')\,=\,{1\over (2\pi)^{10}} \int d^{10}K \;\tilde{G}(K)\,e^{iK\cdot (X-X')}\,\,. 
\ee

\noindent Now, from (\ref{0.3}), (\ref{0.9}), (\ref{0.10}) and (\ref{0.12}) we obtain that,

\be
\label{0.13}
(\Box +\lambda^2\Box_\theta- m^2)\,\int\,{d^{10}K\over (2\pi)^{10}}\,\tilde{G} (K)\,e^{iK\cdot (x-x')}
\,=\,\int\,{{d^{10} x}\over (2\pi)^{10}}\,e^{iK\cdot (x-x')}
\ee
giving the solution for $\tilde{G}(K)$ as
\be
\label{0.14}
\tilde{G} (K)\,=\,-\,{1\over {K^2\,+\,m^2}}
\ee
where, from (\ref{0.8}), $K^2 = K_{(1)\mu}\,K_{(1)}^\mu\,+\,{\lambda^2 \over 2} K_{(2)\mu\nu}\,K_{(2)}^{\mu\nu}$.

Substituting (\ref{0.14}) in (\ref{0.12}) we obtain
\be\label{0.17}
G(x-x',\theta-\theta')\,=\,{1\over (2\pi)^{10}}\int d^{9} K\,\int d\,K^0 {1\over {(K^0)^2\,-\,\omega^2}}\,e^{iK\cdot (x-x')}
\ee
where the ``frequency" in the $(x+\theta)$ space is defined to be
\be
\label{0.18}
\omega\,=\,\omega(\vec{K}_{(1)},K_{(2)})\,=\,\sqrt{\vec{K}_{(1)}\cdot \vec{K}_{(1)}\,+\,{\lambda^2\over 2}K_{(2)\mu\nu}\,K_{(2)}^{\mu\nu}-m^2}
\ee
which can be understood as the dispersion relation in this $D=4+6$ space.  We can see also,
from (\ref{0.17}), that there are two poles $K^0 = \pm\;\omega$ in this framework. Of course
 we can construct an analogous solution for 
$\phi^*_{J} (x,\theta)$.

In general, the poles of the Green's function can be interpreted as masses for the stable particles described by the theory.  We can see directly from equation (\ref{0.18}) that the plane waves in the $(x+\theta)$ space
establish the interaction between the currents in this space and have energy given by $\omega(\vec{K}_{(1)},K_{(2)})$
since $\omega^2=\vec{K}_{(1)}^2+{\lambda^2 \over 2}K_{(2)}^2\,+m^2=K^2_{(1,2)}+m^2$, where
$
K^2_{(1,2)}\,=\,\vec{K}_{(1)}^2+{\lambda^2 \over 2}K_{(2)}^2\,\,.
$
So, one can say that the plane waves that mediate the interaction describe the propagation of  particles in a $x+\theta$ space-time with a mass 
equal to $m$.
We ask if we can use the Cauchy residue theorem in this new space to investigate the contributions of the poles in (\ref{0.17}).  
Accordingly to the point described in section {\bf 3},  we can assume that the Hamiltonian is positive definite and it is directly related to the hypothesis that 
 $K^2_{(1,2)} =-m^2 <0$. However if  the observables are constrained to a four dimensional space-time, due to some kind of compactification, the physical mass can have contributions from the noncommutative sector. This point is left for a forthcoming work \cite{aa}, when we will consider the Fock space structure of the theory and possibles schemes for compactification.

For completness, let us note that substituting (\ref{0.4}) and (\ref{0.17}) into (\ref{0.0}), we arrive at the effective action

\ba
\label{0.32}
S_{eff}\,=\,-\,\int\,d^4 x\, d^6 \theta\, d^4 x'\, d^6 \theta'\, J^* (X)\,\int \frac{d^9 K}{(2\pi)^{10}} \int\,dK^0
\frac{1}{(K^{0})^2\,-\,\omega^2\,+\,i\varepsilon}\,e^{iK\cdot(X-X')}\,J(X')\,\,, \nonumber \\
\mbox{}
\ea
which could be obtainned, in a functional formalism, after integrating out the fields.

\section{Conclusions and perspectives}

In this work we have considered complex scalar fields using a new framework where the object of noncommutativity 
$\theta^{\mu\nu}$ represents independent degrees of freedom.We have started  from a first quantized formalism, where $\theta^{\mu\nu}$ 
and its canonical momentum $\pi_{\mu\nu}$ are considered as operators  living in some Hilbert space.
This structure, which is compatible with the minimal canonical extension of the Doplicher-Fredenhagen-Roberts (DFR) algebra,
 is also invariant under an extended Poincar\'e group of symmetry, but keeping, among others, the usual Casimir invariant operators. After that,
in a second quantized formalism perspective, we succeed in presenting an explicit form for the extended Poincar\'e
generators and the same algebra of the first quantized description has been generated via generalized Heisenberg relations.  This is a fundamental point because the usual Casimir operators for the Poincar\'e group are proven to be kept, permitting to maintain the usual classification scheme for the elementary particles. We also have introduced  source terms in order to construct the general solution for the complex scalar fields fields using the
Green's function technique. The next step in this program is to construct the mode expansion in order to represent the fields in terms of annihilation and creation operators, acting on some Fock space to be properly defined. Also possible compactifications schemes will also be considered. These point are under study and will published elsewhere \cite{aa}.

\section{Acknowledgments}

EMCA would like to thank FAPERJ (Funda\c{c}\~ao de Amparo \`a Pesquisa do Estado do Rio de 
Janeiro) and CNPq (Conselho Nacional de Desenvolvimento Cient\'{\i}fico e Tecnol\'ogico),
Brazilian research agencies, for partial financial support.


\begin{thebibliography}{30}
\bibitem{NCQM} In Refs.   \cite{Amorim1} and     \cite{Amorim4} it is possible to
find a large amount of references concerning noncommutative quantum mechanics.
\end{thebibliography}
\end{document}